\documentclass[fleqn,usenatbib]{mnras}

\usepackage{newtxtext,newtxmath}
\usepackage[T1]{fontenc}
\allowdisplaybreaks
\usepackage{graphicx}	% Including figure files
\usepackage{amsmath}	% Advanced maths commands
\title[]{Weibel-dominated quasi-perpendicular shock: hybrid simulations and in-situ observations}

\author[]{
J.A. Kropotina,$^{1}$\thanks{E-mail: juliett.k@gmail.com}, A.A. Petrukovich$^{2}$, O.M. Chugunova$^{2}$, A.M. Bykov$^{1}$
\\
$^{1}$Ioffe Institute\\
$^{2}$Space Research Institute, Russian Academy of Sciences, Moscow, Russia\\
}

\date{Accepted 2023 July 4. Received 2023 July 4; in original form 2023 April 27}

\pubyear{2023}

\begin{document}
\label{firstpage}
\pagerange{\pageref{firstpage}--\pageref{lastpage}}
\maketitle

\begin{abstract}
We directly compare hybrid kinetic simulations and in-situ observations of a high Mach number high-beta shock in the Solar wind. We launch virtual probes to demonstrate that the model quantitatively reproduces the observations. The observed wave properties are caused by the ion Weibel instability in the shock foot. Parameters of reflected ions in the shock foot are extracted from simulations, and their coordinate dependencies are linearly approximated. These approximations could be used in analytical models. Due to strong magnetic variations at ramp the reflected ions density can be locally very high (nearly that of the incoming flow), which makes favourable conditions for the instability.
\end{abstract}

%Don't make up new ones.

%%%%%%%%%%%%%%%%%%%%%%%%%%%%%%%%%%%%%%%%%%%%%%%%%%

%%%%%%%%%%%%%%%%% BODY OF PAPER %%%%%%%%%%%%%%%%%%

\section{Introduction}

Collisionless shocks propagating through low magnetized plasma appear in various astrophysical objects. The synchrotron radiation of gamma-ray burst afterglows is likely associated with energetic particles and magnetic fields, produced by the Weibel instability mediated shocks driven by relativistic outflows in the low-magnetized circumburst  medium \citep[see e.g.][]{1999ApJ...526..697M,2006ApJ...641..978M,2019PhRvL.123c5101L}. The  spectacular merger and accretion events in the clusters of galaxies are accompanied by the observed large scale shocks which are propagating through a hot intercluster plasma \citep[see e.g.][]{2021A&A...651A..41C,2023A&A...670A.156C, mv07,B19}. The shock Alfv`en Mach numbers and the ratio of thermal to magnetic pressure (i.e. the plasma parameter $\beta$) can be large in the intercluster medium. Also, the magnetic field in the cold expanding supernova ejecta is expected to be extremely low, if one assumes that it is the field of progenitor star scaled according to the magnetic flux conservation law \citep{Ellison2005, Telezhinsky2012}. Hence, the reverse shocks  observed in supernova remnants \citep[see ][for a review]{2018SSRv..214...28R} are likely to be unmagnetized.   

Numerical models and laser plasma experiments \citep{2015NatPh..11..173H,Park2015,Marcowith16} indicate that in the environments initially negligible magnetic fields can substantially grow due to the ion Weibel instability (IWI) which appears during the interaction of incoming and reflected flows \citep{Chang1990, Burgess2016}. The electron Weibel instability arises first and quickly thermalizes electrons. After that the much slower IWI instability comes into play and forms a collisionless shock with a strong electromagnetic turbulence. As a result the magnetic energy density can reach about 1--2~\% of the upstream bulk kinetic energy density \citep{Kato2008}. These magnetic fields not only shape collisionless shocks but are also favorable for magnetic reconnection and subsequent particle acceleration \citep{Bohdan2020}. Thus low-magnetized Weibel-mediated shocks might be a site of electrons pre-acceleration and injection into the first order Fermi acceleration. Meanwhile strong magnetic fields near shock transition increase the  momentum which particles need to enter the Fermi process. So the summary impact of the IWI on the Fermi acceleration efficiency is still an open question. Understanding the microstructure and properties of Weibel-mediated shocks is required to solve this problem. 

Weibel-mediated shocks have been extensively studied by means of kinetic simulations \citep{Kato2008, Kato2010, Spitkovsky2008, Bohdan2020}. Particle-in-cell (PIC) codes used for the simulations are highly resource-intensive because they operate on electron scales and must resolve the Debye length to avoid a nonphysical heating. As far as collisionless shocks form on much greater ion scales, some tricks are routinely used to artificially bring electron scales closer to the ion ones and reduce the computation time. Those tricks include reducing of the proton-to-electron mass ratio and increasing the upstream temperature, both increasing the ratio of the Debye radius to the ion inertial length. The upstream flow velocity (in the shock front reference frame) $V_{sh}$ must be increased correspondingly to keep the sonic Mach number. For this reason PIC simulations usually deal with relativistic or subrelativistic shocks (at least $V_{sh} \sim 0.1c$, where $c$ is the velocity of light). 

In the nonrelativisic case the magnetisation parameter can be estimated as $\sigma \equiv [B_0^2/8\pi]/[n_{0} (m_i + m_e )V^2/2]$, where $B_0$, $n_0$ and $V$ are a far upstream magnetic field, number density and flow velocity; $m_i$ and $m_e$ are proton and electron masses. It was proposed in \cite{Kato2008} that $\sigma$ must be lower than $10^{-4}$ for a shock to be Weibel-dominated. This makes doubtful the presence of Weibel-dominated shocks in older than 1000~yr supernova remnants. Meanwhile in the laser plasma experiment of \cite{Park2015} Weibel structures were found in shock with $\sigma \sim 10^{-3}$. In the solar wind $\sigma$ are even higher (at least $\sim 10^{-2}$). 
Moreover, in \cite{Burgess2016} Weibel structures were obtained in hybrid simulations of low Mach number low-beta shocks with sigma about 0.1. This points out that shocks can be Weibel-dominated in moderately magnetized regime. However, the authors pointed out that hybrid codes have some limitations, and thus their results should be confronted with observations or PIC simulations. They also proposed that the Cluster and MMS spacecraft are capable of resolving Weibel structures.

Near-Earth observations with spacecraft provide unique possibility to study collisionless shock structure
and dynamics in-situ, measuring electric and magnetic field, as well as electron and ion distribution functions. However, low-magnetized (high-$\beta$) conditions are not frequent in solar wind plasmas. Among a very few early direct observations of the low magnetized shocks in the solar wind \citep[e.g.][]{Formisano75, Winterhalter88}, ISEE 1 and ISEE 2 spacecraft revealed important details of the high $\beta$ terrestrial bow shock structure \citep{Farris92}. The large separation between the spacecraft (over 2500 km) allowed simultaneous upstream and downstream solar wind plasma measurements for a period of about 8 min. Large amplitude magnetic field and density fluctuations were measured and a  hot dense field-aligned ion beam escaping from the downstream region of the shock was detected. The beam was associated by \citet{Farris92} with short periodic magnetic holes detected in front of the bow shock. Recent studies of high $\beta$ shocks with MMS, Cluster and Geotail spacecrafts were reviewed by \cite{Petrukovich2021}.
Often the transition region of a high-beta shock contained quasi-periodic linearly polarised pulsations, most probably associated with the IWI \citep{Sundberg2017, Petrukovich2019AnGeo, Petrukovich2021}. 
 But the relation between observed quasi-periodic oscillations and shock structure was never studied in details. 

In this paper we build a hybrid kinetic model of a nonrelativistic high-beta shock observed by MMS and directly compare it with the observations. In our model we study the growth of the magnetic variance in the foot region and find it consistent with the predictions of the kinetic linear theory for the IWI. Also we launch a virtual probe to study the nature of the observed quasiperiodic oscillations. We conclude that the observed  nonrelativistic ($V \sim 400$ km/s) quasiperpendicular shock is formed due to the IWI and has a typical structure with normal-aligned filaments of density and magnetic field. Weibel structures are non-propagating in the plasma reference frame, but they are convected supersonically along the shock surface. This happens because the mean flow velocity along shock surface is substantial in the foot region occupied by reflected ions. Hence waves minima and maxima come across the slowly moving spacecraft and lead to the observed pulsations. 

The quantitative agreement of our hybrid kinetic model and in-situ observations, as well as qualitative agreement with \cite{Burgess2016} allows to verify that hybrid codes are capable of reproducing Weibel-dominated shocks. We also determine the properties of the reflected ions beam responsible for the development of the instability. 

The paper is organized as follows: in Section \ref{sec:theory} we provide the linear theory of the IWI; in Section \ref{sec:observations} we describe an observed event in the Solar wind; in Section \ref{sec:simulations} we introduce our kinetic numerical model and in Section \ref{sec:results} we discuss the simulated shock properties and compare them with the observed ones. The discussion and conclusions are given in Sections \ref{sec:discussion} and \ref{sec:conclusions} respectively.

\section{Theory}

\label{sec:theory}
The transverse
electromagnetic \citet{Weibel_59} instability is widely discussed for a long time in the modeling of collective processes in plasma with anisotropic particle distributions, both in the laboratory plasma installations  \citep[e.g.][]{Weibel_PiC_1971,1972PhFl...15..317D,2004PhRvS...7k4801D} and in the space environment  \citep[see e.g.][]{Balogh, 2011A&ARv..19...42B,2015SSRv..191..519S,Marcowith16,2017SSRv..207..319P, 2023PhPl...30c0901T}. 
The ion beam Weibel instability in a cold unmagnetized cross-field ion beam moving relative to the  static cold magnetized electrons was considered by \cite{Chang1990}. Besides the well-known modified two-stream and lower-hybrid drift instabilities they found a purely growing electromagnetic mode which they called the IWI. Their approach was generalized in \cite{Park2015, Burgess2016} for the case of two opposite cold unmagnetized cross-field ion beams. In the center of mass reference frame the growth rate is given by 
\begin{equation}
\label{eq:gamhydro}
\Gamma^2 = \frac{k^2 n_c n_b (V_c - V_b)^2}{(nc + n_b)^2(1 + k^2c^2/\omega_{pi}^2)},
\end{equation}
where $V$ and $n$ are a flow velocity and number density of both ion populations, $k$ is a wavenumber, $\omega_{pi}$ is the ion plasma frequency. Here the subscript $c$ denotes the denser core and the subscript $b$ --- the fainter beam (note, however, that the expression is symmetric, so the subscripts can be exchanged). The growth rate is independent on $B_0$ and becomes asymptotic to $|V_c - V_b|\sqrt{n_c n_b}/(n_c + n_b)$ for $k \gg \omega_{pi}/c$. The wavevector is perpendicular to the beams.

\cite{Kato2010} studied the IWI kinetically taking the parameters from their PIC simulation. In case when a magnetic field and a wavevector are along $z$, and both beams lie in the $x-y$ plane the dispersion equation reads as
\begin{equation}
    \label{eq:detl}
    \det \Lambda = 0,
\end{equation}
where
\begin{align}
\notag \Lambda_{xx} = 1 - \left(\frac{kc}{\omega}\right)^2 + \frac{1}{2}\left(\frac {\omega_{pe}}{\omega}\right)^2 \xi_0[Z(\xi_1) + Z(\xi_{-1})] + \\ + \sum_s \left(\alpha_s 
+ 2\left(\frac{V_{x,s}}{V_{T,s}}\right)^2(1 + \alpha_s)\right)\left(\frac {\omega_{ps}}{\omega}\right)^2, \\
\notag
\Lambda_{yy} = 1 - \left(\frac{kc}{\omega}\right)^2 + \frac{1}{2}\left(\frac {\omega_{pe}}{\omega}\right)^2 \xi_0[Z(\xi_1) + Z(\xi_{-1})] + \\ + \sum_s \left(\alpha_s + 2\left(\frac{V_{y,s}}{V_{T,s}}\right)^2(1 + \alpha_s)\right)\left(\frac {\omega_{ps}}{\omega}\right)^2,\\
\notag \Lambda_{zz} = 1 + 2\left(\frac{\omega_{pe}}{k V_{T,e}}\right)^2[1 + \xi_0Z(\xi_0)] + \\ + 2\sum_s \left (\frac{\omega_{ps}}{k V_{T,s}}\right)^2(1 + \alpha_s),\\
\notag \Lambda_{xy} = \frac{i}{2} \left(\frac{\omega_{pe}}{\omega}\right)^2\xi_0[Z(\xi_1) - Z(\xi_{-1})] + \\ + 2\sum_s\left(\frac{\omega_{ps}}{\omega}\right)^2 \frac{V_{x,s}}{V_{T,s}}\frac{V_{y,s}}{V_{T,s}}(1 + \alpha_s),\\
\notag \Lambda_{yx} = -\frac{i}{2} \left(\frac{\omega_{pe}}{\omega}\right)^2\xi_0[Z(\xi_1) - Z(\xi_{-1})] + \\ + 2\sum_s\left(\frac{\omega_{ps}}{\omega}\right)^2 \frac{V_{x,s}}{V_{T,s}}\frac{V_{y,s}}{V_{T,s}}(1 + \alpha_s),\\
\Lambda_{xz} = \Lambda_{zx} = 2\sum_s\left(\frac{\omega_{ps}}{\omega}\right)^2 \frac{V_{x,s}}{V_{T,s}}\frac{\omega}{kV_{T,s}}(1 + \alpha_s),
\\ \Lambda_{yz} = \Lambda_{zy} = 2\sum_s\left(\frac{\omega_{ps}}{\omega}\right)^2 \frac{V_{y,s}}{V_{T,s}}\frac{\omega}{kV_{T,s}}(1 + \alpha_s), \\
    \xi_n = \frac{\omega - n\Omega_{e}}{kV_{T,e}}, \quad \alpha_s = \left(\frac{\omega}{kV_{T,s}}\right)Z\left(\frac{\omega}{kV_{T,s}}\right), \\ \Omega_{e} = -\frac{eB}{mc}, \quad \omega_{ps} = \sqrt{\frac{4\pi n_s e_s^2}{m_s}}, \quad V_{T,s} = \sqrt{\frac{2k_B T_s}{m_s}}, \\
    Z(\xi) \equiv \pi^{-\frac{1}{2}}\int_{-\infty}^{\infty}\frac{e^{-z^2}}{z - \xi}dz.
\end{align}
Here $\omega$ is a complex frequency, and the summation is only over ion sorts, i.~e. for $s = b, c$ (beam, core).

This dispersion equation will be solved numerically in section \ref{sec:increment} with parameters taken from our simulations. The solution includes a purely growing mode  which corresponds to the kinetic IWI. The increment is typically much lower than \eqref{eq:gamhydro}.

\section{Observations}

\label{sec:observations}
For  the analysis we used measurements of NASA Magnetospheric Multiscale (MMS) project from  magnetic field (FGM) \cite{Russell16} and plasma (FPI) \cite{Pollock16} experiments. In order to directly compare simulations and observations we chose the bow shock crossing by MMS spacecraft on November 25, 2017 (see \cite{Petrukovich2021}). This is a high-beta strong collisionless shock in a region with an ambient magnetic field $B_0$ as low as $0.9$~nT. The Alfv\'{e}n Mach number in the shock rest frame is $M_a \approx 60$ and the shock inclination angle is $\theta \approx 65^\circ$. with a total ion number density $n_i \approx 9$~cm$^{-3}$. The protons' temperature is $T_p \approx 1.1$~eV, and the electrons' temperature is $T_e \approx 13.4$~eV.  

Shocks with such parameters are rare in solar wind, only about 30 well
documented cases for $\beta>30$ were found in the observations by modern 
spacecraft \citep{Petrukovich2021}. About half of these cases have rather rich  internal structure with extended variations, similar to the event presented here, while the other have the appearance closer to a more standard shock (a single magnetic field and density jump). It should be noted that this difference in appearance is not due to the angle between shock normal
and upstream magnetic field (parallel shocks are known to have more extended variations than perpendicular ones). Most of considered shocks have this angle larger than 45$^\circ$.

\begin{figure}
    \centering
    \includegraphics[width = \linewidth]{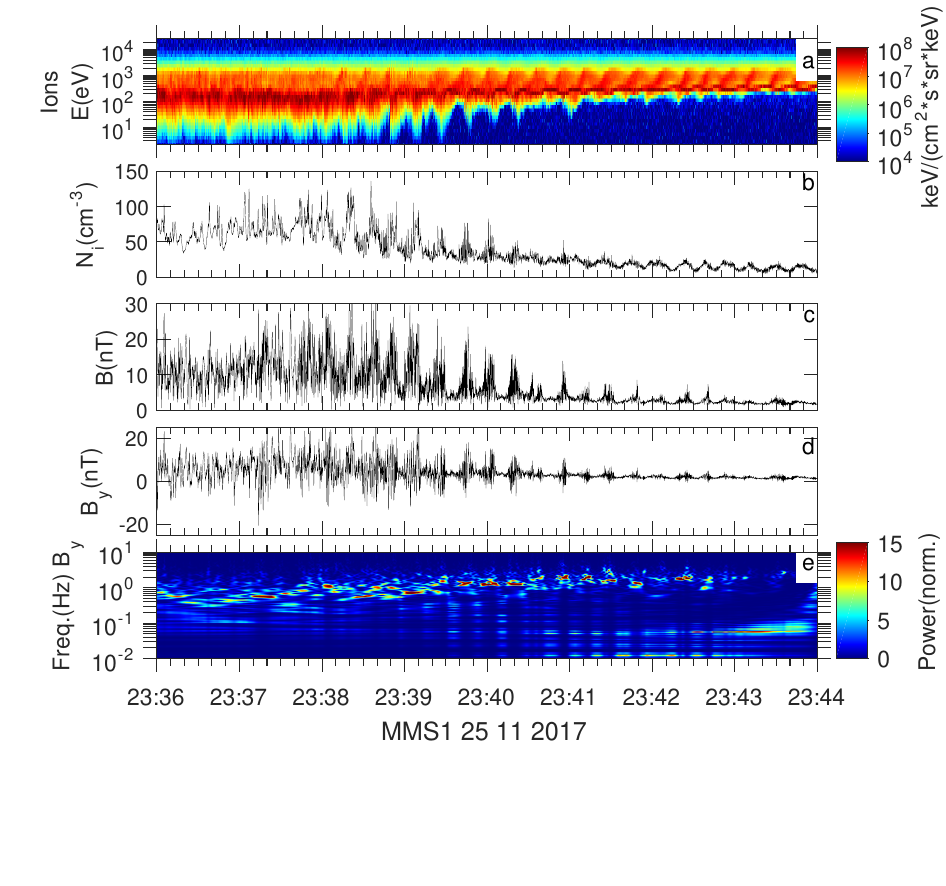}
    \caption{MMS observations of the bow shock crossing at November 25, 2017. (a) Ion omnidirectional
spectrogram; (b) ion number density; (c) magnetic field magnitude; (d) $B_y$ magnetic
field; (e) wavelet dynamic spectrum of magnetic field $B_y$.}
    \label{fig:observations}
\end{figure}

Though the shock crossing takes some minutes (Fig. \ref{fig:observations}), the physical width of the shock layer is only about one proton cyclotron radius in the (very low) upstream magnetic field. The spacecraft gradually crosses the shock from downstream to upstream and observes relatively stable picture of periodically ($\sim$15 s) emerging activations, gradually thermalizing the solar wind ion flow.  Each activation, in turn, consists of high-amplitude magnetic variations with a period about 1 s, coupled with pulses of a downstream-like plasma flow. Sometimes these density and magnetic field peaks are higher than the downstream averaged plasma density and field values. Between the pulses more upstream-like flow is observed with a substantial fraction of reflected and accelerated ions. 

Available observations with four closely separated spacecraft allow to determine the wavelength of 1-sec pulsations of about 150 km as well as the propagation velocity and direction. These waves have linear polarisation and are almost standing in the plasma rest frame, consistent with the expectation for the Weibel mode. Later on we compare these values with those obtained in simulations. 

\section{Simulations}
\label{sec:simulations}

We modeled a shock with parameters taken from the observations by means of the hybrid code ``Maximus'' \citep{Kropotina2019, Kropotina2021ApJ}. We used a 3d cartesian grid sized $L_x \times L_y \times L_z = 2500 \times 150 \times 150$~cells, each cell $0.1 l_i^3$, where $l_i$ is the proton inertial length. The shock was launched via the rigid piston method, when a super-Alfv\'{e}nic flow with a bulk velocity $V_x = -45 V_a$ was reflected from a conductive wall at $x = 0$. This resulted in a formation of a shock front moving in the positive $x$ direction with $V_{f} \approx 15.3 V_a$. Thus in the shock front frame $M_a = 60.3$. The initial magnetic field lay in the $x-z$ plane at an angle $\theta = 65^\circ$ to the shock normal. Average values for Helium content (4~\% He(+2) by number) and temperature (He(+2) is four times hotter than protons)
were used in the model. Thus the total mass density was $\rho_0 \approx 1.7\cdot10^{-23}$~g/cm$^3$. Plasma parameters for all particle sorts were $\beta_p \equiv 8\pi 0.96 n_i T_p / B_0^2 = 4.8$, $\beta_{He} = 8\pi 0.04 n_i 4T_p / B_0^2 = 0.8$, $\beta_e = 8\pi 1.04 n_i T_e / B_0^2 = 62.4$, and total $\beta = \beta_p + \beta_{He} + \beta_e = 68$.
Electrons were treated as neutralizing massless fluid with the adiabatic equation of state and standard adiabatic index $\Gamma = 5/3$. 

It should be noted that hybrid codes cannot capture electron kinetics, thus the highest-frequency modes might be modeled incorrectly. However, our model is highly resource-intensive even within the hybrid approach. Meanwhile the same simulation box size seems to be unreachable in frames of full PIC modeling, especially with realistic electron-to-proton mass ratios (which in turn might affect the results). For this reason we chose the hybrid approach. The comparison with observations will validate our method at least in the sense of reproducing wave directions, amplitudes, spectra, and polarisation.

In the hybrid code all quantities are normalized, i.e. a magnetic field and a mass density are measured in $B_0$ and $\rho_0$, lengths are measured in $l_i = c/\omega_{pi} \approx 2.3 \cdot 10^7 n_0^{-0.5}$~cm, times --- in the inverse proton gyrofrequencies $\Omega^{-1} = m_p c /e B_0 \approx 11.6$~s, velocities --- in the Alfv\'{e}n velocities $V_a = B_0 / \sqrt{4\pi \rho_0} \approx 5.8$~km/s, temperatures are given in energy units $m_p V_a^2 \approx 0.4$~eV. To make units self-consistent we took as $n_0$ the number density of a pure proton plasma with the same $\rho_0$ (see \cite{Matthews94}). Thus for the proton-helium plasma $n_0$ was slightly greater than the electron number density $n_e$. All simulation parameters are listed in Table \ref{tab:scales}.

\begin{table}
    \centering
    \begin{tabular}{|c|c|c|c|}
    \hline
         $B_0$ & 0.9 nT & $T_p$ & 1.1~eV\\
         $n_i$ & 9 cm$^{-3}$ & $T_{He}$ & 4.4~eV\\
         $n_0$ & 10 cm$^{-3}$ & $T_e$ & 13.4~eV \\
         $l_i$ & 68 km & $L_x$ & 2500 cells \\
         $\Omega_{ci}^{-1}$ & 11.6 s & $L_y$ & 150 cells\\
         $V_a$ &  5.8 km / s & $L_z$ & 150 cells\\
         $M_a$ & 60.3 & cell size & 0.1 $l_i^3$\\
         $\theta$ & $65^\circ$ \\
             \hline
    \end{tabular}
    \caption{Simulation parameters}
    \label{tab:scales}
\end{table}

\section{Results}
\label{sec:results} 
\subsection{Shock structure}

\begin{figure}
	\includegraphics[width=\columnwidth]{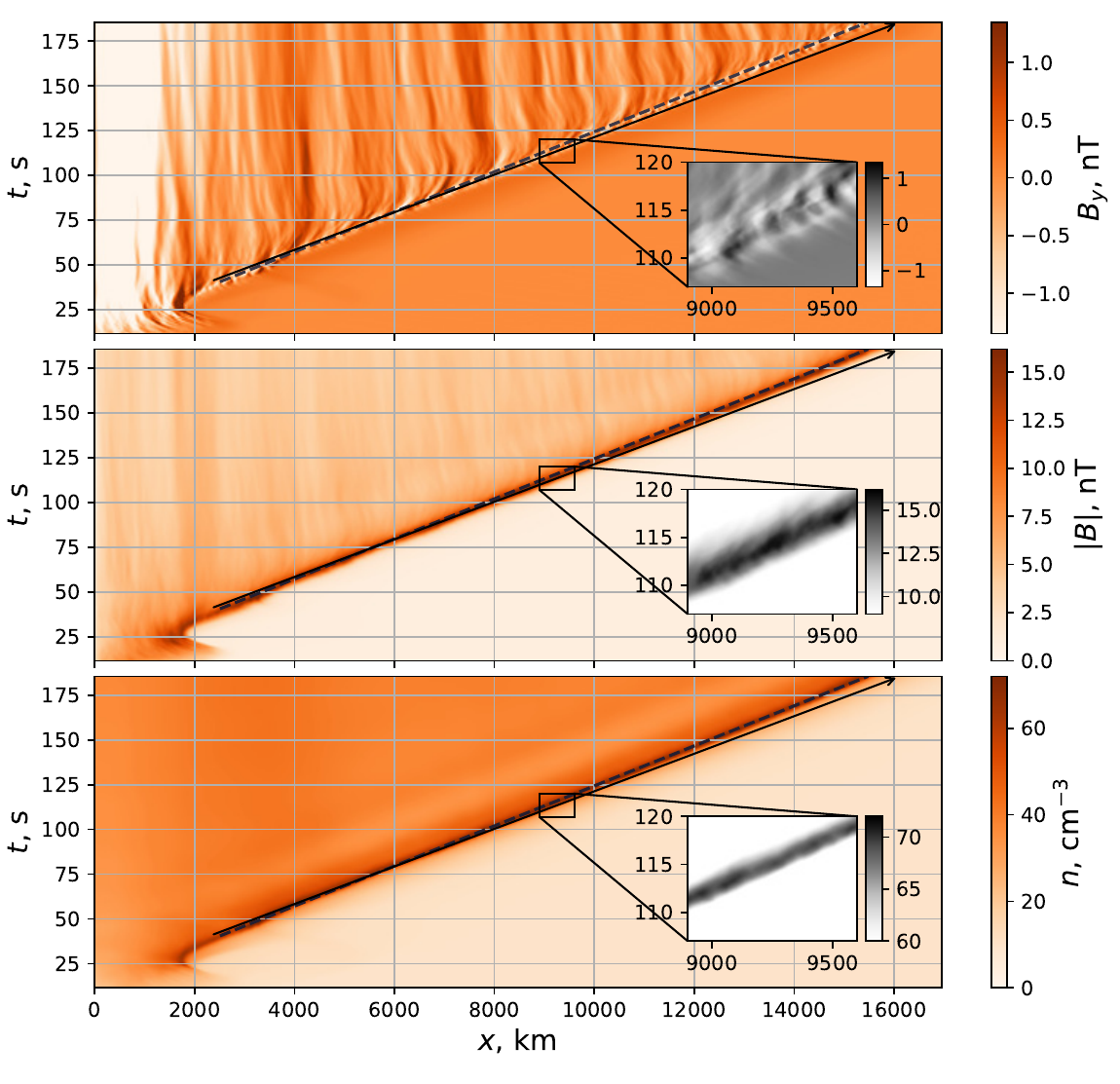}
    \caption{Shock dynamics. From top to bottom: cross-section averaged transverse magnetic field $B_y$, magnetic field magnitude and density. An approximate front position is marked by dashed gray lines. Black arrows represent a trajectory of a virtual probe. The insets represent the closer view of the front.} 
    \label{fig:evolution}
\end{figure}

The shock temporal evolution is color-coded in Fig. \ref{fig:evolution}. The front is formed at $t \approx 30$ s ($\sim 3 \Omega_{ci}^{-1}$) and propagates with a nearly uniform velocity $V_{f} \approx 15.3 V_a \sim 89$~km/s (its trajectory is shown by dashed gray lines in all panels). To mimic the MMS shock crossing four virtual probes were launched. They were located in vertexes of a right tetrahedron with an edge equal to 0.3~$l_i\approx 22$~km. These virtual spacecraft moved along the shock normal from the downstream to the upstream with $V_{p,x} = 16.3 V_a$ ($\approx 1 V_a$ in the front rest frame) and measured magnetic field and plasma parameters on their way. Their trajectories are shown in Fig. \ref{fig:evolution} by a black arrow (the distance between the probes is not resolved). The resulting temporal profiles are discussed in section \ref{sec:probes}. We checked that transverse probe movement with $V_{p,y} = V_{p,z} = 0.3 V_a$ didn't introduce any differences.

\begin{figure*}
	\includegraphics[width=\textwidth]{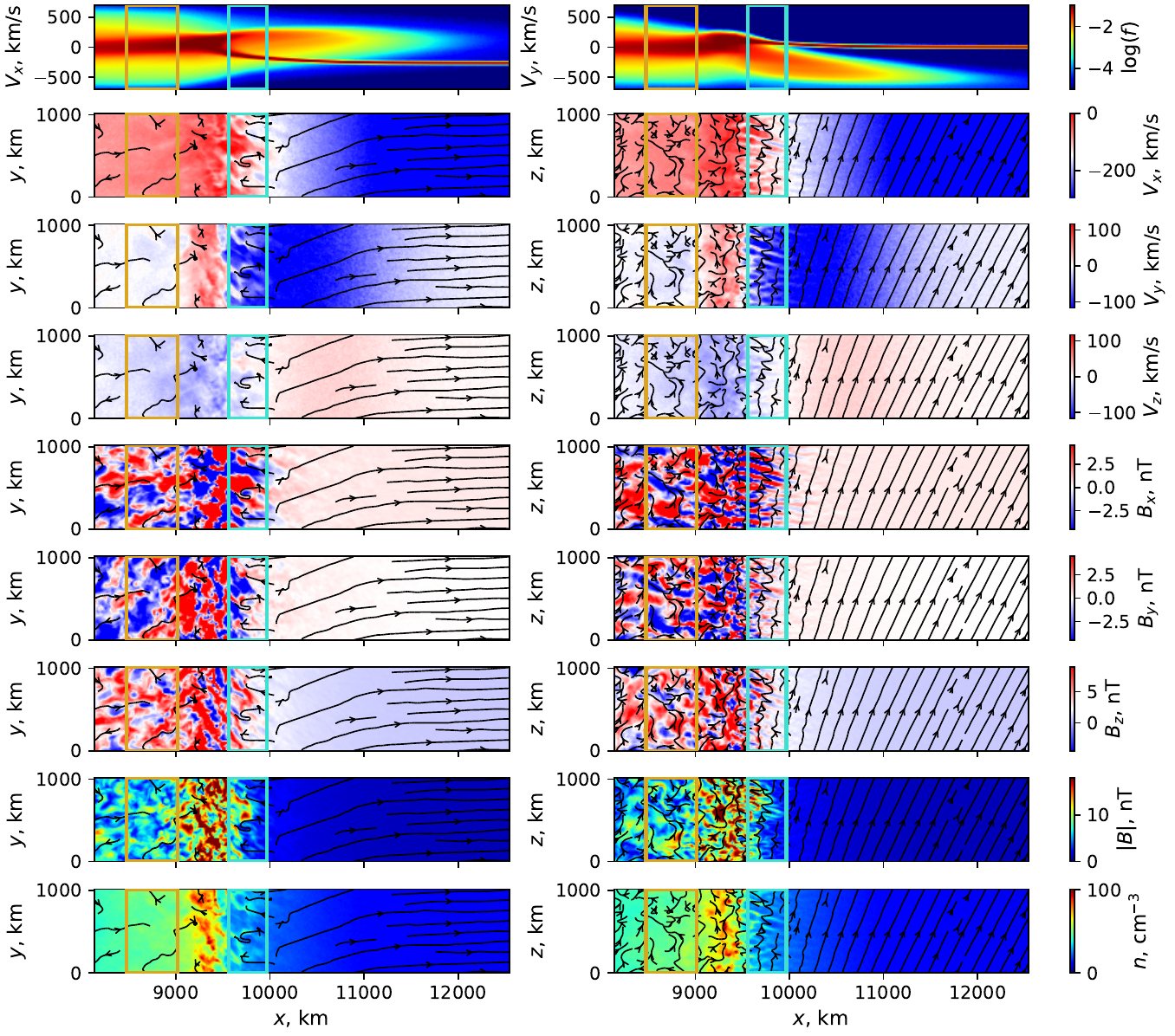}
    \caption{Shock structure. Top panel: $x-V_x$ and $x-V_y$ protons phase spaces  averaged over $y$ and $z$. Other panels (from top to bottom): velocity, magnetic field and number density maps in two projections. $V_x$ is given in the front rest frame.  Colored rectangles mark downstream and upstream zones where spectral analyses were made. Magnetic field lines are superimposed in black.}
    \label{fig:maps}
\end{figure*}

\begin{figure}

	\includegraphics[width=\columnwidth]{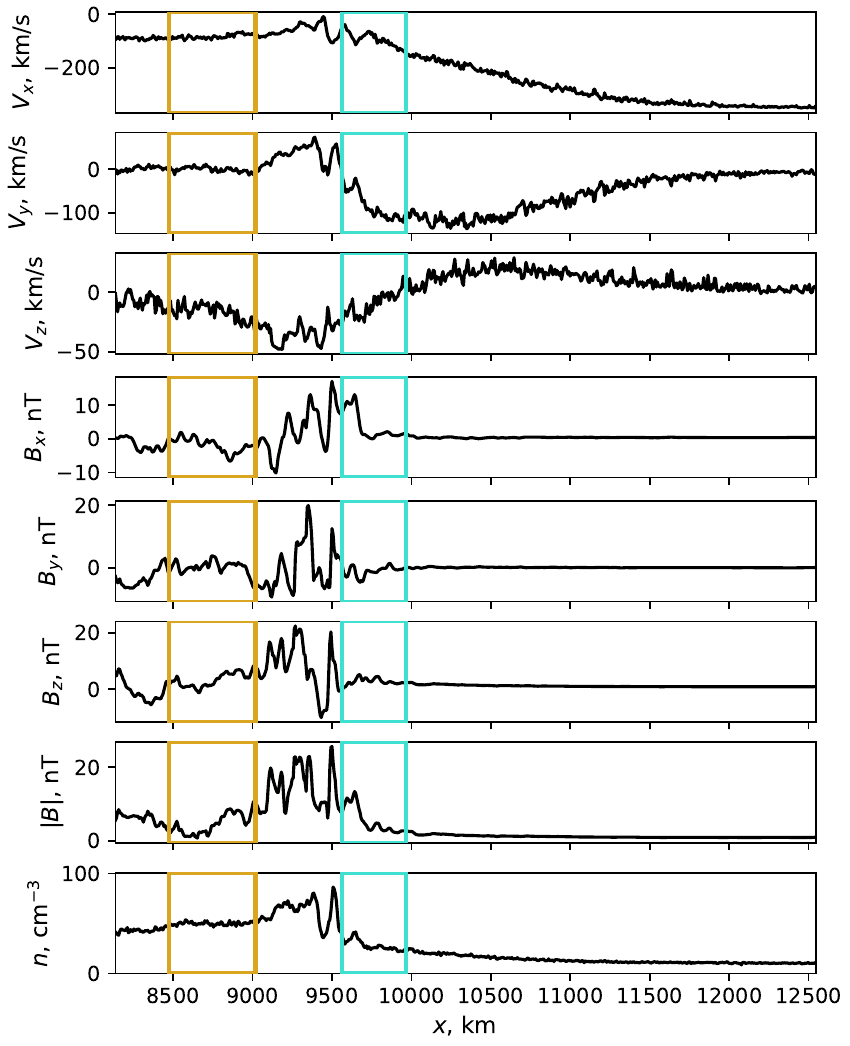}
    \caption{Flow velocity, magnetic field and density along the shock normal $y = z = 0$ at $t = 10 \Omega_{ci}^{-1}$. Selected regions are the same as in Fig. \ref{fig:maps}}
    \label{fig:profiles}
\end{figure}

\begin{figure}
	\includegraphics[width=\columnwidth]{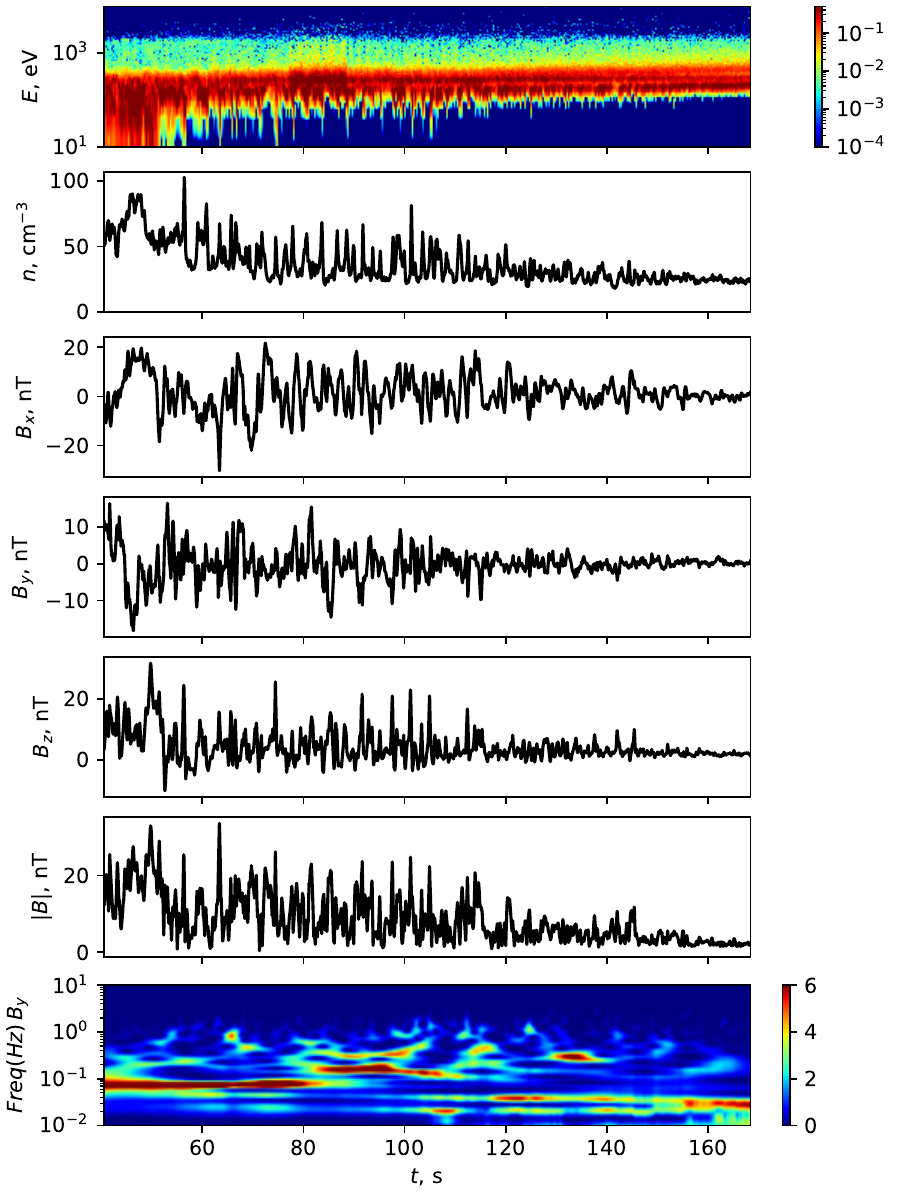}
    \caption{Overview of the Probe~1 shock crossing. From top to bottom: protons phase space $x-E$; ion number density; magnetic fields; wavelet dynamic spectrum of $B_y$.}
    \label{fig:probe}
\end{figure}

The shock structure at $t = 10\Omega^{-1} \approx 2$  min after initialisation is shown in Figures \ref{fig:maps} and \ref{fig:profiles}. The shock front has a complicated filamentary structure, and its surface is highly corrugated (rippled). Ion phase spaces in the top panels of Fig.~\ref{fig:maps} show that the shock foot extends over several thousands kilometers. Specularly reflected ions gyrate in the upstream magnetic field and induce a cross-field current along $x$ and $y$. This is auspicious for the IWI. A detailed inspection of $V_x$ and $n$ maps revealed that high density filaments correspond to higher negative velocities along normal. This indicates that density enhancements in the foot are not due spatial variations of the number of shock-reflected ions. On the contrary density variations appear further in the foot and are convected towards the shock front due to the plasma bulk flow.

In the shock coplanarity plane (Fig.~\ref{fig:maps}, right column) thin structures (filaments) are seen in the foot region. They make a small angle with a shock normal and grow rapidly towards the front.  The overall picture resemble those seen in simulations of \cite{Kato2008, Kato2010}, who concluded that such structures appear due to the IWI. Structures seen in both planes are also very alike those discussed in \cite{Burgess2016}. They identify narrow Weibel filaments with width close to $2l_i$ in the coplanarity plane and somewhat wider oblique 'tongues' in the perpendicular plane which they call 'the AIC-like ripples'. 

Magnetic fluctuations associated with these filaments are very strong near the front. In the downstream (yellow rectangle) and in the close foot region (cyan rectangle) their amplitude reaches $10 B_0$, and between them, at ramp, it is twice as large. Thus the magnetic energy density reaches about 2~\% of the bulk kinetic energy. Number density peaks reach about $10 \rho_0$. 

\subsection{Virtual probes}
\label{sec:probes}

Instantaneous shock profiles along normal (Fig. \ref{fig:profiles}) are much smoother than those seen by MMS (Fig. \ref{fig:observations}). Hence we cannot directly compare observations with simulations if we ignore relative motion of the shock and spacecraft.  In Fig. \ref {fig:probe} we show what observes one of the virtual probes, which starts at $t = 3.5 \Omega^{-1}$ at $x = 37.5 l_i$. The picture is very alike the observations (see Fig. \ref{fig:observations}) and severely differs from the instantaneous profiles (Fig. \ref{fig:profiles}). The reason is that the plasma moves across the probes much faster than the probes move across the shock. Moreover, the plasma transverse movements are highly oscillatory. We found that along probes' ways $V_y$ varies between $-30 V_a$ and $15 V_a$, and $V_z$ --- between $-15 V_a$ and $5 V_a$. 

To study how the observed oscillations change while the probe is moving across shock we performed the Morlet wavelet transform of $B_y$ projection and found that the picture qualitatively resembles MMS observations (Cf. the bottoms panels of Fig. \ref{fig:observations} and Fig. \ref{fig:probe}).
The main difference between the model and observations is the absence
in the model of the clear-cut bunching of 1-Hz oscillations in the 
$\sim$15 sec 'packets', though some enhancements are observed in the frequency spectra at about 0.1 Hz (bottom panel of Fig. \ref{fig:probe}).

It should be noted that our virtual spacecraft crossed the shock much faster than the real ones (two minutes vs five). It was done in order to show the whole transition with reasonable computational efforts. However to check the impact of probes velocity we also launched several probes starting ar different points and moving with $V_{p,x} = 15.5 V_a$ and $V_{p,x} = 15.7 V_a$ (i.~e. $0.2 V_a$ and $0.4 V_a$ relatively to the front). 

\begin{figure}
	\includegraphics[width=\columnwidth]{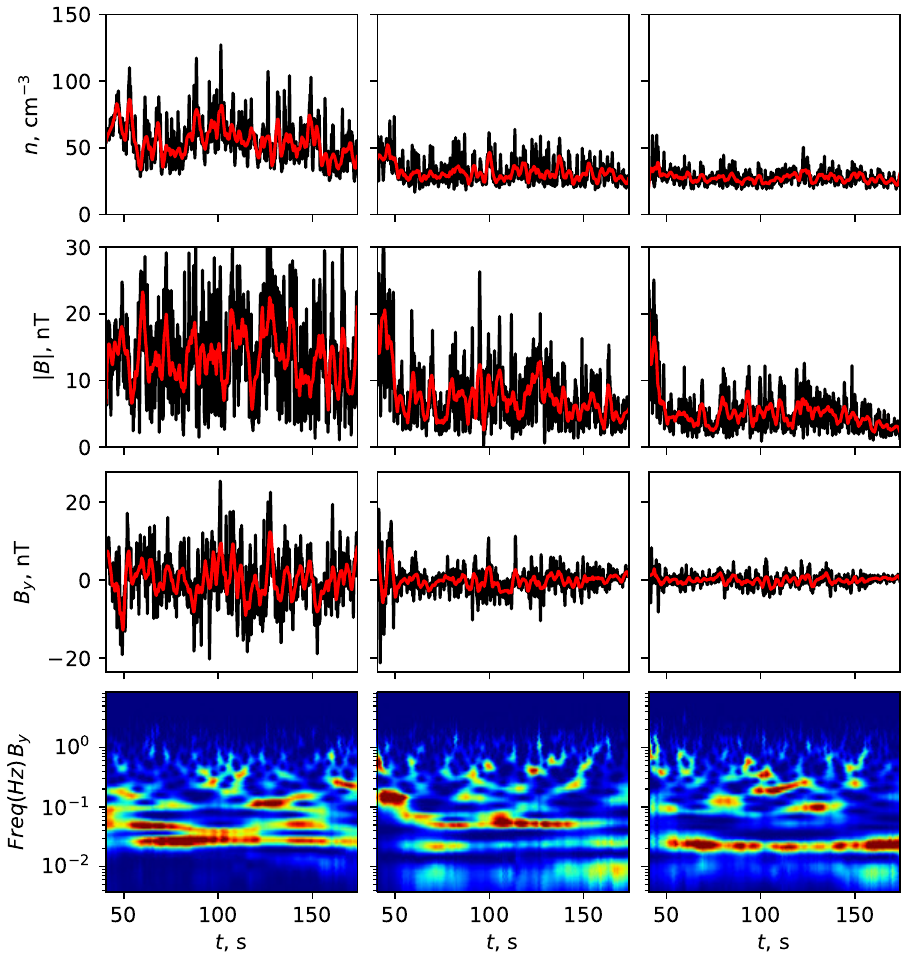}
    \caption{Measurements of probes moving with $0.4V_a$ relatively to the shock front. Left column -- a probe starting at $x = 38 l_i$; middle column -- at $x = 41 l_i$; right column -- at $x = 42.5 l_i$. Red lines show the same data smoothed by a Savitzky-Golay filter.}
    \label{fig:slowprobes}
\end{figure}

Fig. \ref{fig:slowprobes} shows the results of three probes starting at $x = 38 l_i$, $41 l_i$ and $42.5 l_i$ and moving with $V_{p,x} = 15.7 V_a$. They should cross the shock in $\sim5$~minutes, just as the real spacecraft. Black lines show the probes measurements, and red ones correspond to the smoothed data. 

The modeled data seems to be more noisy than due to limited number of particles per cell. Also the highest frequencies might be affected by the grid resolution and the lack of electrons kinetics. Hence the modeled curves do not ideally reproduce the observed ones. However rather prominent wave packets appear in the case of slower probes, which stay longer in each region. More upstream-like regions with lower density and magnetic fields alternate with more downstream-like ones. The wave packets are less clearly separated from each other than in observations. This probably indicates that the observed shock is more variable. One of the reasons might be that the longest waves are restricted by simulation box sizes. 

Overall, it is possible that prolonged shock crossings like the presented one are observed due to extremely low proper shock speed (of the order of km/s). The detailed discussion of this issue is beyond the scope of this paper. Here we concentrate on comparison of the observed and simulated structures.

We compare the plasma wave properties observed at the near-Earth shock, with those in the recordings of the virtual probes with $V_{p,x} = 16.3 V_a$ (Table 2). Four probes in observations and simulations allow to determine not only the temporal sequence of measured parameters, but also determine spatial gradient (hence, wavevector) on a scale of separation. In both cases magnetic oscillations were linearly polarised (the maximum variance eigenvalue is at least 4-5 times larger, than the medium variance one) and
the wavevector was close to the local magnetic field ($\theta_{kB}$ < 40$^o$).
Dominating frequency (measured as frequency of the peak in the spectrum)
in observations is larger: 1.25 Hz vs 0.38 and 0.8 Hz (two equivalent peaks are present in the simulation interval). 

The doppler shift in frequency is in both cases close to the observed frequency, which means that the
waves are standing in the rest frame and are purely convected with the plasma flow.  Observed and simulated wavelengths are about 100-200 km. 
Concluding, we consider the properties of magnetic oscillations in observation and simulation as very similar in their principal characteristics.

\begin{table}[t]
\caption{Wave analysis data for P\#2 and P\#3.}
\begin{tabular}{l c c}
\hline
Parameter & observation & simulation \\
& 23:40:15--23:40:21.5 UT &  97.1--102.7 s\\
observed frequency, Hz & 1.25 & 0.37, 0.8 \\
eigenvalues  & 4.72,    5.78,   46.2  & 11.9 22.47 63.99 \\
observed wave speed, km/s & 173 & 85\\
wavelength, km & 138 & 230, 106 \\
Doppler shift, Hz & 1.3 & 0.37, 0.81\\
$\theta_{kB}$  & 35$^o$ & 37$^o$ \\
\hline
\end{tabular}
\end{table}

 There are oscillations with frequencies between $10^{-1}$ and 1~Hz with quasiperiodic enhancements and frequency growth towards upstream.

\subsection{Shock dynamics}

Supercritical collisionless shocks are known to be quasi-stationary. The transition appears due to partial reflection of the incoming ions and reforms quasi-periodically. Shock reformation is a topic of great interest, actively investigated by means of numerical models and in-situ observations \citep[see, e.~g.][]{Turner2021, Yang2020, Johlander2022}. There are two mechanisms of this process: (I) the accumulation of reflected ions in a foot until their density becomes comparable to that at ramp, and (II) the front interaction with waves convected by the upstream flow \citep{Marcowith16}. The insets of Fig. \ref{fig:evolution} show some signatures of the first type reformation: the front velocity and the cross-section averaged magnetic field at the overshoot slightly vary with time. However, field variations are relatively weak, and a density profile is nearly stationary. So the ``classical'' picture of shock reformation did not reveal in this case. The more thorough investigation of this problem is beyond the scope of this paper.

Meanwhile waves generated by the IWI and convected by the flow substantially contribute to the shock nonstationarity as well. To demonstrate this we made a real-time movie of the probe with $V_{p,x} = 15.7 V_a$ recordings together with its movement through the shock. The movie is available online in the supplementary materials. Fig. \ref{fig:probemaps} shows one frame of this video. In the upper row color maps of $B_y$, $n$ and $V_x$ are given. The velocity is in the front rest frame. The probe position is marked by a white triangle, and its recordings are shown in the bottom row. The red line corresponds to the data measured until the current moment, and the blue one -- to the future recordings. 

From the color maps we can see that the front is highly corrugated. In the movie all these structures move both towards the shock and across it, leading to a lively structure and oscillatory probe measurements. We chose the moment when the probe is near a density peak, which corresponds to a higher negative $V_x$. Such regions appear where transverse magnetic fields are low and the upstream plasma easily penetrate downstream. These ``paths of least resistance'' are surrounded by regions with higher $B_\perp$, where hotter reflected ions lead to pressure increase. This pressure compresses colder ``paths of least resistance'' up to nearly downstream density. So thin dense filaments appear, clearly visible in the upper middle panel of Fig. \ref{fig:probemaps}. 

\begin{figure}
	\includegraphics[width=\columnwidth]{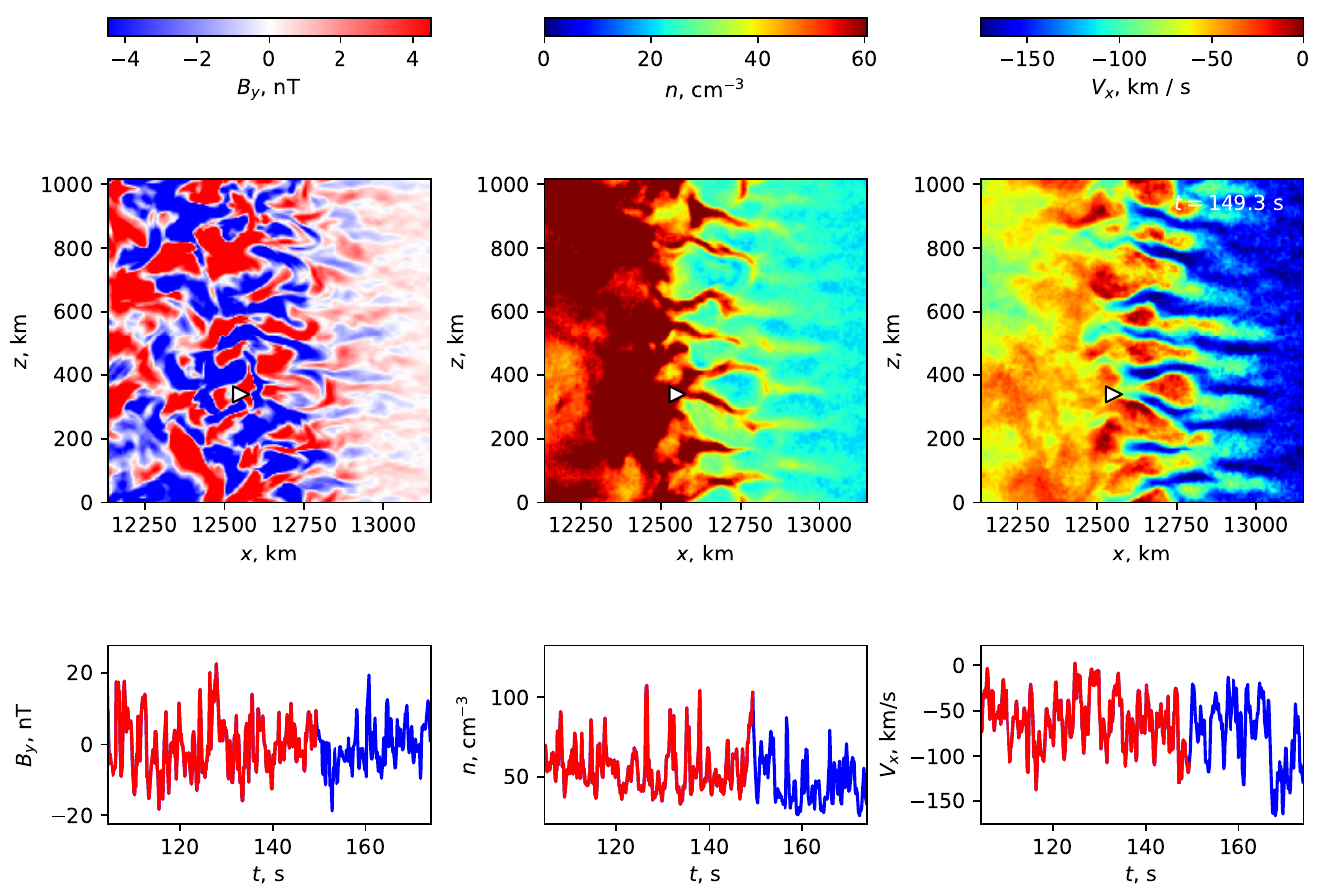}
    \caption{A snapshot of the shock transition with a virtual probe marked by a white triangle. Upper row: color maps of $B_y$, $n$ and $V_x$, lower row: the probe recordings of the same quantities. The current moment is at the conjunction of red (past) and blue (future) lines. The corresponding movie is available online.}
    \label{fig:probemaps}
\end{figure}

\subsection{Spectral analyses}

\label{sec:spectra}
\begin{figure*}
	\includegraphics[width=\textwidth]{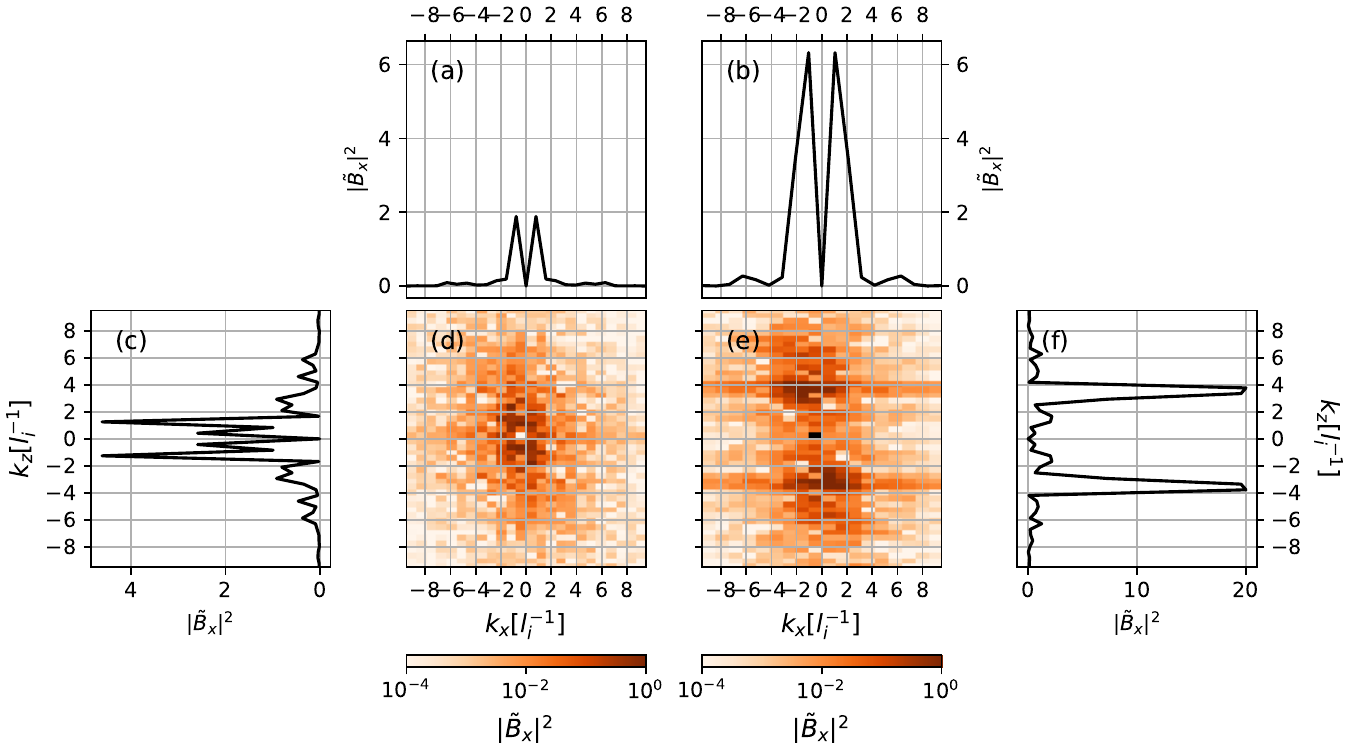}
    \caption{Panel (d): $k_x - k_z$ maps of spectral power in $B_x$ fluctuations in the downstream region (inside the yellow rectangle in Fig. \ref{fig:maps}); panel (e): the same in the close foot region (cyan rectangle in Fig. \ref{fig:maps}); panels (a) and (b): 1d spectral power of $B_x(x)$ at $z = 0$ in the corresponding regions;  panels (c) and (e): 1d spectral power of $B_x(z)$ at the left edges of the corresponding regions.}
    \label{fig:kxkz}
\end{figure*}

\begin{figure*}
	\includegraphics[width=\textwidth]{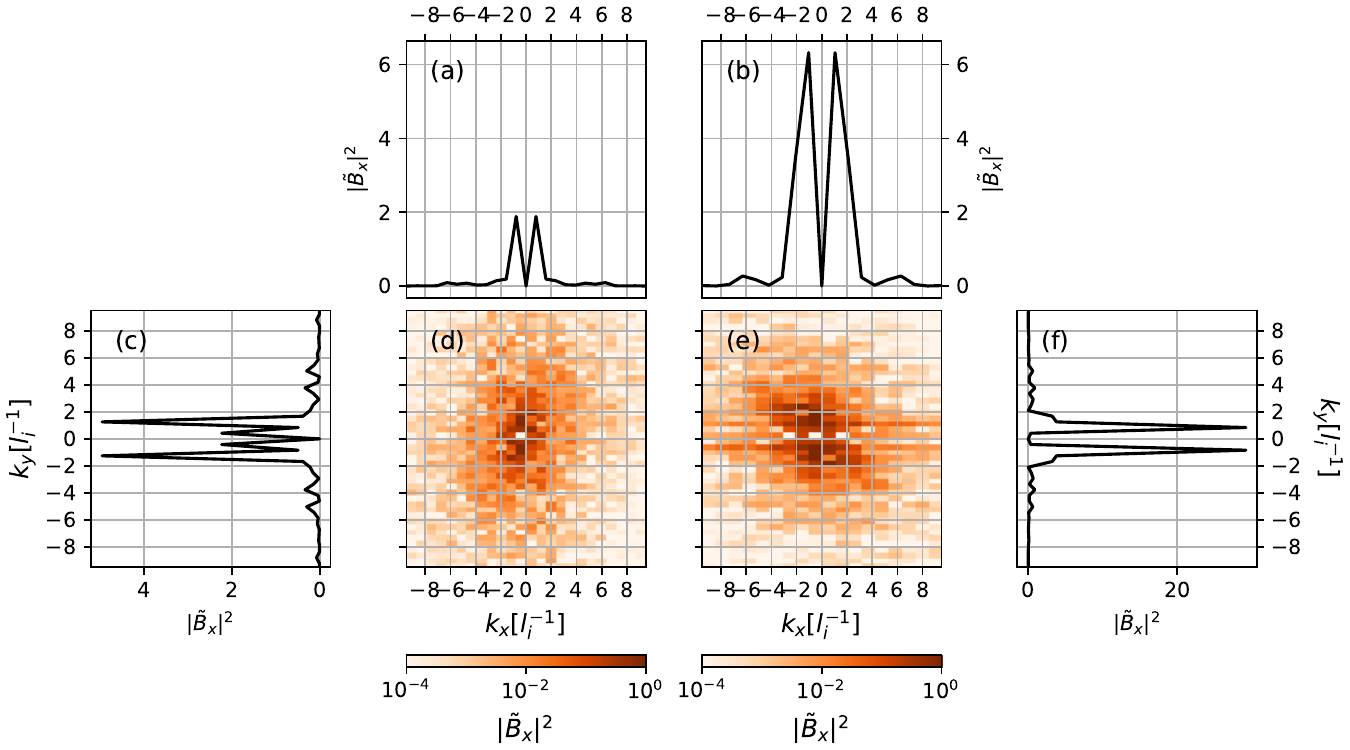}
    \caption{The same as in Fig. \ref{fig:kxkz}, but for $k_x-k_y$}
    \label{fig:kxky}
\end{figure*}

To better understand the structure of the shock we made spectral analyses of magnetic field fluctuations in  regions marked by yellow and cyan rectangles in Figures \ref{fig:maps} and \ref{fig:profiles}. Figures \ref{fig:kxkz} and \ref{fig:kxky} show $B_x$ spectral power density in $k_x - k_z$ and $k_x - k_y$ planes in the close downstream region marked by a yellow rectangle, and in the close upstream region marked by a cyan rectangle. The spectral power density is $|\tilde B_{x}|^2$, where $\tilde B_{x}$ is a discrete Fourier transform of $B_x$. From 1d and 2d spectra it can be seen that in the close upstream region $k_z \approx 3.5 l_i^{-1}$ and $k_x \approx k_y \approx 1 l_i^{-1}$, which corresponds to $\lambda \sim 2 l_i \sim 140$~km. In the close downstream wavelengths are larger. We also checked that in the close upstream $B_x$ fluctuations are stronger than those of $B_{y,z}$, while in the close downstream the spectral power in all three magnetic field projections is comparable.

The upstream wavevector direction and waves polarisation in the coplanarity plane $x-z$ agree with those expected for the IWI, i.~e. $\textbf{k}$ is along the mean magnetic field and, hence, is nearly perpendicular to the cross-field beam in the $x-y$ plane (see top panels of Fig. \ref{fig:maps}). 

\subsection{Growth rate analyses}

To demonstrate that the shock transition is governed by the IWI we directly compare the growth rate and spectral properties with linear predictions. We expect a zero frequency mode, thus if we move towards the shock front with the flow, the waves amplitude grows as 
\begin{equation}
\label{eq:growth}
 b_0 \exp\left(\int\Gamma(x(t)) dt \right),
\end{equation}
where  $b_0$ is its initial value and $\Gamma$ is an increment.

\begin{figure}
	\includegraphics[width=\columnwidth]{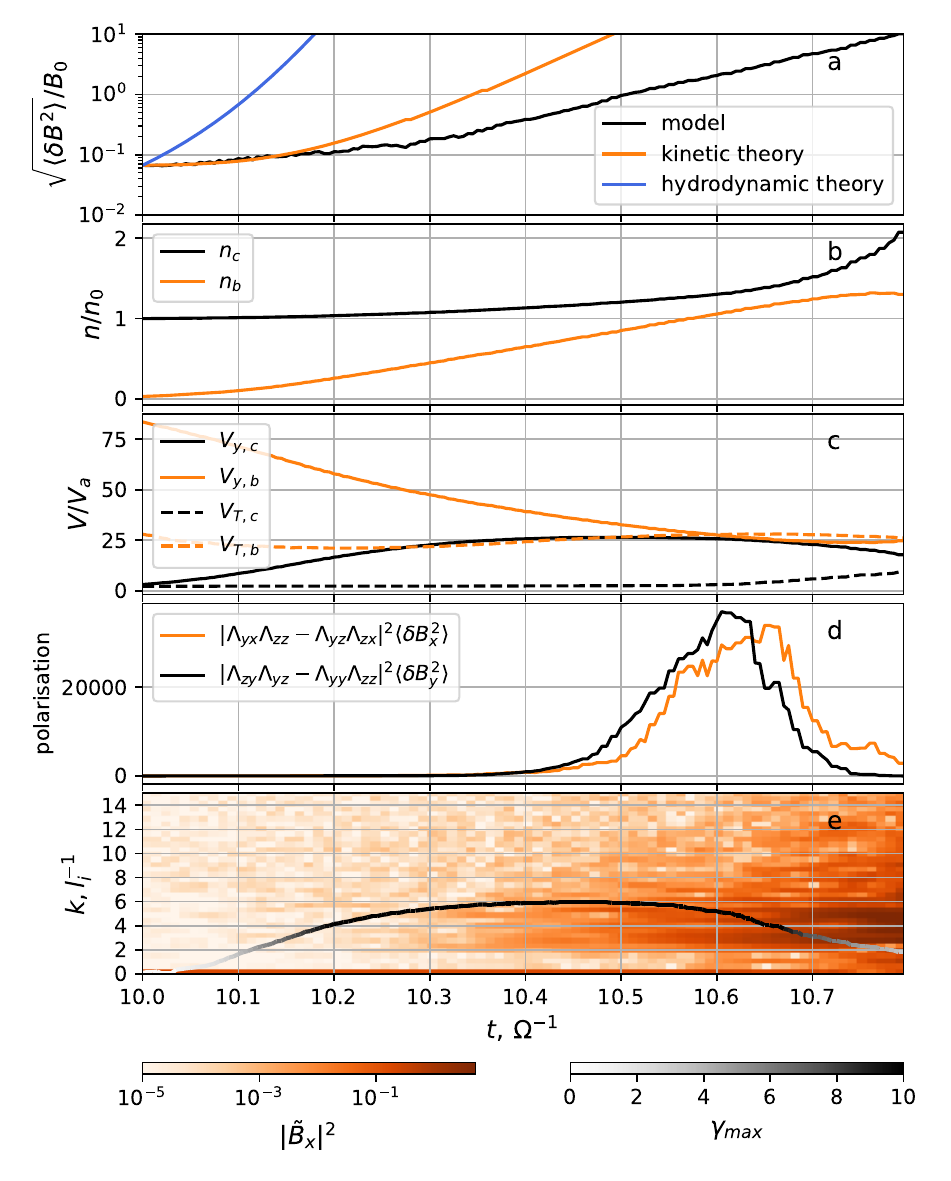}
    \caption{Growth rate analyses: (a) the simulated growth of magnetic fluctuations amplitude towards shock front compared with the predictions of hydrodynamic and kinetic linear theories; (b) beam and core number densities; (c) beam and core thermal velocities and flow velocities along $y$ in the center of mass rest frame; (d) polarisation analyses (see text); (e) color-coded $B_x$ fourier spectrum with overlaid linear prediction of $k_{max}$ (the curve color codes the maximal predicted growth rate).}
    \label{fig:growth}
\end{figure}

To check this we studied the evolution of magnetic field fluctuations amplitude towards shock front. We calculated it as a standard deviation over a transverse slice embedded in the upstream flow. The result is shown in Fig. \ref{fig:growth}, a.

We integrated the $x-V_x$ and $x-V_y$ phase spaces (see the top panels of Fig. \ref{fig:maps}) to estimate velocity, density and temperature of the core and beam. We arbitrarily placed a boundary between them at $V_x = -15 V_a\approx -90$~km/s and $V_y = -5 V_a \approx -30$~km/s. We also checked that the result is nearly the same for boundaries at $V_x = 0$ and $V_y = -10 V_a$. Core and beam densities, as well as their thermal and flow velocities in the center of mass rest frame are shown in the panels (b) and (c) of Fig. \ref{fig:growth}.

Knowing physical parameters at each point we could find hydrodynamic and kinetic increments and maximal wavenumbers from \eqref{eq:gamhydro} and \eqref{eq:detl} respectively. It should be noted that the kinetic approach suggests treating protons and helium ions separately. However, we considered unmagnetized ions, so only their plasma frequencies and thermal velocities are important. The latter are equal in our model because ions temperatures are mass-proportional. The ratio of He(+2) and proton plasma frequencies depends only on their number densities, as if they were both protons. So we solved \eqref{eq:detl} considering a pure proton plasma. 

Knowing $\Gamma$ and taking initial $b_0$ from simulations we directly compared simulated and theoretical growth using \eqref{eq:growth}. Theoretical amplitudes are shown in the panel (a) of Fig. \ref{fig:growth} by orange and blue curves, and the predicted wavenumber is superposed on the actual spectrum in the panel (e). The curve color codes the corresponding increment. We also checked that the real frequency found from the dispersion equation was zero.

The linear kinetic theory can also predict waves polarisation. In respect that $\Lambda_{ij}E_j = 0$, and $c \mathbf{k} \times \mathbf{E} = \omega \delta\mathbf{B}$, where $\mathbf{E}$ is an electric field and $\delta\mathbf{B}$ is a magnetic field variation, and $\mathbf{k}$ is along $z$, we find that $$(\Lambda_{yx}\Lambda_{zz} - \Lambda_{yz}\Lambda_{zx})\delta B_y = (\Lambda_{zy}\Lambda_{yz} - \Lambda_{yy}\Lambda_{zz})\delta B_x.$$  In panel (d) of Fig. \ref{fig:growth} we compare $|\Lambda_{yx}\Lambda_{zz} - \Lambda_{yz}\Lambda_{zx}|^2\langle\delta B_y^2\rangle$ and $|\Lambda_{zy}\Lambda_{yz} - \Lambda_{yy}\Lambda_{zz}|^2\langle\delta B_x^2\rangle$, where the magnetic variance is taken from simulations, and $\Lambda_{ij}$ is calculated from the beam properties. The curves do not perfectly coincide, but they resemble each other even in a highly nonlinear regime.

It can be seen that the hydrodynamic increment is far too large, but the simulated growth rate is reasonably explained by the kinetic linear theory until the wave amplitude approaches about $0.1 B_0$. After that the system gradually enters a nonlinear regime, and the predicted growth rate outplays the actual one. Note also that \eqref{eq:detl} was obtained for a uniform medium, and the actual increment may differ due to strong gradients. The predicted wavenumbers are slightly higher than the simulated spectral maxima, but the simulated spectrum is rather broad, so the agreement is satisfactory. The polarisation properties of the IWI are also well reproduced. So we can conclude, that the IWI governs the shock transition.

\label{sec:increment}

It should be noted that the investigated shock has a sound Mach number $M_s$ as low as 7. However, linear analyses in \cite{Nishigai2021} indicated that a shock must have both $M_a$ and $M_s$ as great as $\sim 20-40$ to be Weibel-dominated. The authors argued that the instability behaves Weibel-like if the growth rate is much greater than the ion cyclotron frequency. For an Alfven Mach number of 60 and a sound Mach number of 7, Fig. 3 of \cite{Nishigai2021} predicts the growth rate which is comparable and slightly larger than the ion cyclotron frequency. In our simulations it locally reaches $\sim10$.

In \cite{Nishigai2021} the reflected ions in the foot  were parameterised as a ring distribution with a number density about $0.2 n_0$, the radius of the ring equal to the upstream flow velocity, and thermal spread equal to the one of the upstream flow. From Fig. \ref{fig:growth} it can be seen that the actual quantities strongly vary along the shock normal. Closer to the shock the reflected ions density nearly reaches that of the incoming flow. In the regions where the density is about $0.2 n_0$ the beam velocity is greater than the upstream one. That is why the IWI growth rate exceeds that predicted by \cite{Nishigai2021}. 

Simple analytical models like that in \cite{Nishigai2021} are a powerfool tool to scan a wide range of parameters with minimal computational efforts. So it is useful to precise them with parameterisation of ions distributions based on numerical models. To make the first step in this direction we approximated the reflected ion density and flow velocity in the simulated shock foot by simple linear functions of coordinate:
$$n_b/n_0 = 2.0(1.0 + (x - x_{sh}) / R_g),$$
$$|V_{b,y} - V_{c,y}|/V_a = 0.66M_a(1.0 + (x - x_{sh}) / R_g),$$
where $R_g$ is the effective particle gyroradius in the foot. It appeared to be equal to $0.4M_a l_i$ in our case because the transverse magnetic field is greater than that far upstream. A thermal velocity of the beam varied only slightly and was close to $25 V_a$ (an order higher than that of the core). The beam velocity along $x$ could not be approximated linearly, but it quickly reached a relatively stable value $|V_{b,x} - V_{c,x}|/V_a \approx M_a$. The difference between $V_{b,x}$ and $V_{b,y}$ is due to the shock drift acceleration by a motional electric field along $y$ axis \cite{Sagdeev1966}. As a result $V_{b,y}$ eventually exceeds $V_{b,x}$, and $B_x$ variation becomes stronger than that of $B_y$.

\section{discussion}
\label{sec:discussion}

Near Earth spacecraft plasma observations afford unique possibility to sample in-situ such important astrophysical phenomena as collisionless shocks. A rich variety of shock structures was discovered, depending on basic plasma constants and geometry (Mach number, plasma $\beta$, magnetic field direction, etc). However these experiments are essentially limited by number of spacecraft simultaneously available — measurements can be performed only in few points, while the spatial structure at large is only inferred. On the other hand,
numerical modelling affords the full access to spatio-temporal structure of the shock transition. To make calculations to be completed in realistic time, 
simplifications to physical models are usually introduced, which may question the applicability of results. 

In our work, we were able to prove that rather typically observed high $\beta$, high-M$_a$  shock structure  with the developed high-amplitude magnetic fluctuations is well reproduced with our hybrid model with Helium and adiabatic electrons. Consistency is found in the general appearance of the shock transition (Fig.1 and Fig.5), as well as in quantitative characteristics of the dominating plasma wave mode (Tab. 2). 

It is shown that the temporal profiles of the shock crossing depend substantially on the relative velocity of the shock front and the spacecraft probes. Slowly flying probes (as mostly in space experiment) are able to detect the strong temporal variability of the shock front, while high-speed motion results in rather simple almost instantaneous profile cuts (Fig. 4). It is not always possible to determine the spacecraft-shock relative velocity in orbit and possibility of such strong dependence of observations on relative motion should be taken into account.

Of course, observed differences of the shock structure (see discussions in \citet{Petrukovich2019AnGeo,Petrukovich2021} might be due to some differences in shock parameters such as magnetic field angle and Mach number. Some our simulation runs, not shown here, reveal significant variance of shock structure across parameter range and model details, even if all cases are high-$\beta$ shocks. This parametric dependence of shock structure is left for the future studies. 

Yet another advantage offered by simulations, is the ability to access the 3D spatial structure of the transition region in full details. The cuts of simulation box (like Fig. 3) reveal the complicated breathing filament structure with varying  scale in different directions. These filaments move rapidly along the shock front and create the magnetic and plasma variability observed by the probes. The amplitude of these variability is very large, magnetic amplitudes are order of magnitude larger than the background magnetic field. Such variability might provide sites of magnetic reconnection and particle acceleration, though in our case we have not seen it neither in the simulations nor in the observations, probably due to high $\beta$ value.

Plasma properties also principally change across filaments: more sheath-type thermalized and high-density streams interchange with more upstream-type with low density and high percentage of the reflected ions. Close to the ramp the reflected ions density nearly reaches that of the incoming flow. The detailed physics of such complicated shock transition remains to be studied with point-by-point comparison of observations and simulations. It is important for such a study, as it was stated above, that our numerical model is closely compatible with observation in all comparable properties. 

Finally, our results represent one more proof that high-$\beta$ shock transition is dominated with the Weibel-like plasma wave mode. We determine polarisation as well as dispersion characteristics, which coincide in observations and modeling. The complicated spatial structure detected, suggests that such mode needs to be considered in the deeply non-linear regime, practically shaping the process of plasma flow thermalisation. 
Knowing parameters of magnetic variations allows to analyse variants of shock-related particle acceleration and diffusion at such astrophysical objects. To improve analytical models of shock transition we extracted the parameters of the reflected ions distribution from our simulation and found that the beam density and flow velocity could be well approximated by linear functions.

\section{Conclusions}
\label{sec:conclusions}

We demonstrated that hybrid kinetic models can quantitatively reproduce observed properties of strong Weibel-dominated high-beta quasiperpendicular shocks. Hybrid models are much less resource-intensive than PIC ones and do not need high upstream temperatures and subrelativistic flow velocities. This gives a possibility to study a large field of shock parameters and find the conditions when shocks become Weibel-dominated. Strong magnetic variations at ramps of such shocks could prevent particles injection into the first order Fermi acceleration process. On the other hand such variations might cause magnetic reconnection which in turn produces nonthermal particles. So the net impact of the IWI on particle acceleration is still to be determined.  

We also extracted from simulations distributions of reflected ions in the shock foot. This allows to improve existing analytical models of such shocks.

\section*{Acknowledgements}

JK and AB acknowledge the Russian Science Fund grant 21-72-20020,
which supported the plasma numerical modeling presented here. 
Some of the modeling was performed at the Joint Supercomputer Center JSCC RAS and at the “Tornado”
subsystem of the St. Petersburg Polytechnic University supercomputing center.
 AP and OC acknowledge the Russian Science Fund grant 19-12-00313,
which supported the observation analysis and comparison with simulations. Authors are grateful to NASA MMS project team for excellent space project and observations. We are very grateful to the reviewer Dr.~Takanobu Amano, whose fruitful suggestions greatly improved this paper.

%%%%%%%%%%%%%%%%%%%%%%%%%%%%%%%%%%%%%%%%%%%%%%%%%%
\section*{Data Availability}

MMS spacecraft data are open at the NASA CDAWeb data archive \url{https://cdaweb.gsfc.nasa.gov/}.



\bibliographystyle{mnras}
\bibliography{biblio}

%\begin{thebibliography}{99}
%\end{thebibliography}

%%%%%%%%%%%%%%%%%%%%%%%%%%%%%%%%%%%%%%%%%%%%%%%%%%

%%%%%%%%%%%%%%%%% APPENDICES %%%%%%%%%%%%%%%%%%%%%

%%%%%%%%%%%%%%%%%%%%%%%%%%%%%%%%%%%%%%%%%%%%%%%%%%

% Don't change these lines
\bsp	% typesetting comment
\label{lastpage}
\end{document}